\documentclass[10pt, conference]{IEEEtran}

\ifCLASSINFOpdf
  \usepackage[pdftex]{graphicx}
\else
  \usepackage[dvips]{graphicx}
\fi

\usepackage{adjustbox}
\usepackage{array}
\usepackage{subfigure}
\usepackage{url}
\usepackage{pbox}
\let\labelindent\relax
\usepackage{enumitem}
\usepackage{xspace}
\usepackage{enumitem}

\setlist[enumerate,1]{%
  label=\arabic*.,
}

\newlist{inlinelist}{enumerate*}{1}
\setlist*[inlinelist,1]{%
  label=(\arabic*),
}

\begin{document}
%
\title{Flexible In-The-Field Monitoring \\ \large Extended Abstract}


\author{\IEEEauthorblockN{Oscar Cornejo}
\IEEEauthorblockA{
  Department of Informatics, Systems and Communication\\
   University of Milano Bicocca, Milan, Italy\\
  oscar.cornejo@disco.unimib.it\\
  }
}
\maketitle

\begin{abstract}

Fully assessing the robustness of a software application in-house is infeasible, especially considering the huge variety of hardly predictable stimuli, environments, and configurations that applications must handle in the field. For this reason, modern testing and analysis techniques can often process data extracted from the field, such as crash reports and profile data, or can even be executed directly in the field, for instance to diagnose and correct problems. In all these cases, collection, processing, and distribution of field data must be done seamlessly and unobstrusively while users interact with their applications.

To limit the intrusiveness of in-the-field monitoring a common approach is to reduce the amount of collected data (e.g., to rare events and to crash dumps), which, however, may severely affect the effectiveness of the techniques that exploit field data. 

The objective of this Ph.D. thesis is to define solutions for collecting field data in a cost effective way without affecting the quality of the user experience. This result can enable a new range of testing and analysis solutions that extensively exploit field data.
\end{abstract}


\IEEEpeerreviewmaketitle

\section{The Challenge of Collecting Field Data} \label{sec:introduction}

In-field executions can be an invaluable source of information for the developers who need to know how their applications behave when operating in the end-user environment. The collected data might reveal unpredicted usage scenarios, silent faults, inefficiencies, and more in general can provide an accurate picture of the actual behavior of an application.%

The importance of observing the software when running in the field has been already well recognized by industry and academia. For instance, the video streaming company Netflix has reported that its system has grown so much that performing realistic testing in-house is nearly impossible, so it has started testing and collecting data directly in-the-field, using fault-injection and monitoring~\cite{Basiri:Netflix:ISSRE:2016}. Data retrieved from the field have been also exploited 
to profile applications~\cite{Elbaum:Profiling:TSE:2005}, 
debug software~\cite{Clause:FieldFailuresDebugging:ICSE:2007}, 
and reproduce failures~\cite{Jin:BugRedux:ICSE:2012}. 

While just collecting data is simple (it is enough to inject probes into the target system), doing it in a non-intrusive way and without affecting the user experience can be extremely challenging. Note that in this work we focus on applications that contemplate an interaction with the user.%

The impact of monitoring depends on its pervasiveness (e.g., the number of application elements being monitored) and its thoroughness (e.g., the amount of data saved every time an event is collected). While collecting a crash report can be done inexpensively at the time of the crash, capturing field data more extensively, such as collecting coverage data and function calls, is far more expensive~\cite{Orso:DeployedSoftware:FoSER:2010}. Going further, the overhead introduced by massively collecting field data could affect the overall user experience, making the monitoring tasks intrusive and unacceptable for the vast majority of users. 
  
The events that must be collected from the field depend on the \emph{goal} of the monitoring activity, which defines the type of events to be detected (e.g., method calls), and the data that must be reported in the traces (e.g., state information and parameter values). 
Depending on the specific goal, the monitor might need to collect \emph{single} or \emph{sequences of} events. For instance, if the purpose of the monitoring is to calculate branch coverage from field data, each single event, that is the information about the branch taken at a decision point, is useful by itself. While, if the purpose of the monitoring is to calculate path coverage from field data, the information about the branch taken at a single decision point is not useful, since the knowledge about the sequences of all the branches taken for a same execution is required to identify the path that a program executed. Although both cases target the same set of events, the distinction between targeting single events or sequences of events is important. In fact, in the former case a \emph{partial trace}, that is a trace obtained by recording only a subset of the events, carries partial but useful information. While in the latter case, only a \emph{complete trace}, that is a trace obtained by recording all the relevant events produced in an execution, is useful to achieve the monitoring goal. 

When \emph{partial traces} are already useful, there are multiple techniques that can be used to reduce the overhead~\cite{Delgado:TaxonomyFaultMonitoring:TSE:2004}. In particular, 
the set of collected events can be limited arbitrarily~\cite{Elbaum:Profiling:TSE:2005} (Selective Monitoring), 
can be distributed among multiple instances of a same application running on different machines~\cite{Orso:GammaSystem:ISSTA:2002} (Distributive Monitoring), 
and can be determined probabilistically~\cite{Liblit:BugIsolation:SIGPLAN:2003} (Probabilistic Monitoring).

Unfortunately, the useful cases that may produce the most relevant advances in verification and validation techniques require collecting \emph{complete traces}.
Example of goals that require collecting complete traces are monitoring sequences of API calls, system callbacks, and the user activity. 

The problem of cost-effectively monitoring data in the field might be further complicated by the cost of saving the individual events. Most of the techniques that work with field data focus on tracing \emph{simple information}, such as the name of an invoked method, and little has been done about non-intrusively collecting \emph{expensive information}, such as the content of data structures and application states, which may require recording the values of many objects and variables. 

\emph{The objective of this Ph.D. thesis is to study how to satisfy complex monitoring goals by cost-effectively collecting complete traces that may include expensive information for some of the collected events}. To this end, we aim to define a novel class of monitors that can answer to these research questions exploiting the spatial and temporal dimensions.

We will consider the \emph{spatial dimension} to study how to obtain complete traces from multiple partial traces, obtained by selectively distributing the monitoring workload on different user machines. The idea is that a \emph{complete trace} might be feasibly reconstructed from \emph{partial traces} as long as the partial traces are annotated with information that indicates how traces can be merged, we call this kind of annotation \emph{signature data}.

Signature data are dynamic data, such as sequences of events or variable values, that can be recorded and exploited to determine if two different application instances reached a same state at some point of the execution. When a same state is detected at two different points of two distinct partial traces, the traces can be merged at that point, producing new and longer traces representing more extensively the behavior of the monitored software. 

The challenge is finding the proper information that should be reported as signature. Indeed, signature data should not be too expensive to record. At the same time, it should be abstract enough to frequently have the chance of combining traces collected from different instances and different executions, but also concrete enough to perform this operation safely, with little risk of creating infeasible sequences of events. 


Investigating the \emph{temporal dimension} implies finding how to reschedule the monitoring activity so that its overhead can be masked to the user, even when a relevant amount of data must be traced. The intuition is that normally events are collected synchronously as reaction to actions performed by users. However, different users actions may produce or not executions with a noticeable overhead, as discussed in the next section. So, intuitively, the monitor could adaptively anticipate or posticipate some operations, depending on the overhead that has been introduced in the system and the state of the available resources. For example, if enough resources are available, a monitor could save state information in advance under the assumption that this information is likely to stay unchanged until it needs to be saved. If the operation succeeds, the monitoring overhead would be redistributed along the execution 
(e.g., by recording complex objects along the whole execution instead of concentrating the recording of the full object in a single point of the execution) in a way to make it less noticeable to users. 

\section{Research Status and Future Work} \label{sec:status}

During the first year of the Ph.D., we studied how the monitoring overhead may actually impact on the quality of the user experience. 
This is an important aspect because it provides precise indications about what can be collected transparently and what cannot be feasibly done in the field. 
Specifically, we measured the overhead introduced in typical usage scenario by a probe that collects sequences of method calls.

The results show that the amount of overhead that can be tolerated by users is significant~\cite{Cornejo:InTheField:ICSE:2017}. An overhead up to $30\%$ can be hardly recognized by users, while overhead values between $30\%$ and $80\%$ are noticed only moderately. Finally, when the overhead reaches values greater than $80\%$, users can almost always recognize that the execution is abnormally slowing down. These results show that monitors can potentially perform a relevant number of actions in the field without letting users to perceive their presence.

Furthermore, we discovered that the overhead is distributed unevenly across actions: there are actions presenting no overhead, and actions presenting significant overhead, although the same kind of information is collected (sequences of method calls in our case). 

This result suggests that the actions that can be monitored inexpensively could be exploited to anticipate activities that might reduce the overhead introduced in the actions that are more expensive to monitor. 

The next steps of this work includes studying the \emph{spatial dimension} of monitoring by developing a strategy that enables the distribution of the monitoring workload even when sequences of related events must be recorded. The objective is to collect partial traces that can be merged into more extensive, likely complete, traces by exploiting dynamically collected \emph{signature data}.

The strategies to exploit the \emph{spatial} and \emph{temporal dimensions} to reduce the monitoring overhead will be implemented as part of a monitoring framework that will be used to systematically evaluate the approach. 

With this Ph.D. thesis we expect to define new monitoring solutions for the testers and the developers who need to collect an extensive amount of data from the field without affecting the end-user environment.

\vfill
\medskip \begin{small} \emph{Acknowledgments} This work has been partially supported by the H2020 Learn project, which has been funded under the ERC Consolidator Grant 2014 program (ERC Grant Agreement n. 646867).
\end{small}

\vfill

\bibliographystyle{IEEEtran}
\bibliography{biblio}

\end{document}